# *In Silico* Design, Extended Molecular Dynamic Simulations and Binding Energy Calculations for a New Series of Dually Acting Inhibitors against EGFR and HER2

**Marawan Ahmed[1*], Maiada M. Sadek[2], Khaled A. Abouzid[3] and Feng Wang[1]**

[1] eChemistry Laboratory, Faculty of Life and Social Sciences, Swinburne University of Technology, Melbourne, Victoria 3122 Australia.

[2] Pharmaceutical Organic Chemistry, Faculty of Pharmacy, MSA University, 6th October, Egypt.

[3] Pharmaceutical Chemistry Department, Faculty of Pharmacy, Ain Shams University, Cairo 11566, Egypt.

## ABSTRACT

Starting from the lead structure we have identified in our previous works, we are extending our insight understanding of its potential inhibitory effect against both EGFR and HER2 receptors. Herein and using extended molecular dynamic simulations and different scoring techniques, we are providing plausible explanations for the observed inhibitory effect. Also, we are comparing the binding mechanism in addition to the dynamics of binding with two other approved inhibitors against EGFR (Lapatinib) and HER2 (SYR). Based on this information, we are also designing and *in silico* screening new potential inhibitors sharing the same scaffold of the lead structure. We have chosen the best scoring inhibitor for additional *in silico* investigation against both the wild-type and T790M mutant strain of EGFR. It seems that certain substitution pattern guarantees the binding to the conserved water molecule commonly observed with kinase crystal structures. Also, the new inhibitors seem to form a stable interaction with the mutant strain as a direct consequence of their enhanced ability to form additional interactions with binding site residues.

*To whom correspondence should be addressed. E-mail: mmahmed@swin.edu.au,

Tel: 61-3-9214 8785.

# 1. INTRODUCTION

Kinases are known to be a viable target for anti-cancer drug development [1-4]. It's estimated that around 30% of all drug design effort are dedicated for inhibition of kinases [5]. They are the second most important drug targets after GPCR's (G Protein Coupled Receptors) [6]. Kinases are involved in many pathophysiological problems especially cancers where their over-expression can lead to certain types of malignant tumours [7,8]. Several kinase inhibitors are now in the market and many are under clinical trials. Resistance to current kinase inhibitors are emerging and wide-spreading, so, there's an urgent need for the development of more effective analogues that can tackle the problem of resistance [9]. In our two previous studies, we have presented a new series of a quinazoline based Lapatinib analogues with a potential anti-tumour activity and tested them against Epidermal Growth Factor Receptors and Human Epidermal growth factor Receptors 2 tyrosine kinases (EGFR/HER2) [10,11]. Indeed, several members of the new family showed micro-molar inhibitory activities and utilizing new binding patterns [10,11].

It's widely known that ligand binding to kinases is derived mainly by lipophilic vdW interactions [12,13]. In most cases, kinase inhibitors binds to the ATP binding pocket of the kinase domain of the enzyme, and they form several H-bonds with the backbone and side chains of the amino acids present in this region (Hinge region) [14]. Previous studies have shown that for these kinds of receptors including ligand induced polarization by protein environment may be important and ligand binding is best described by QM/MM docking [15,16]. Our previous results have also shown that the Lipophilic-vdW term may be the major factor that derives binding in this new series of inhibitors [10,11]. Other studies have also indicated that the electrostatic effect may be important in certain cases [3].

The ligand-protein binding process is too complex to be described by a single representation of the ligand-protein complex produced as a result of the rigid receptor docking [17]. The various levels of approximations necessary to make docking rapid make scientists sceptical to really believe its results [18]. Proteins are not static idles, they are a very complex, moving and viable machines [19,20]. Introducing protein flexibility is increasingly important to describe the ligand-protein binding especially for targets known to be highly flexible such as kinases [21-23]. Kinases are known to be a very challenging target for molecular modelling, particularly conventional docking. In our previous studies, we have utilized different docking and scoring methods to include some receptor flexibility [10,11]. These methods ranged from rigid docking with vdW radii scaling [17], induced fit docking (IFD) [24,25] and finally molecular dynamics (MD) simulations for some inhibitors against HER2 receptor [10,11].

In the current study, we are extending our insight understanding of the binding of the new series to their targets, EGFR and HER2 utilizing a much longer MD simulations. These kinds of studies are important as certain conformational changes are not accessible at the conventional MD time scale. We are also planning to use our existing knowledge from these simulations in order to design new powerful inhibitors that can inhibit EGFR/HER2 more efficiently. Another motivation is to design members that are less liable to the problem of resistance to current chemotherapeutic agents. As references, we are comparing the new series with the both originally co-crystallized ligands, Lapatinib (EGFR) and SYR (HER2). Also, owing to the observed binding capability of the new series to interact with certain additional residues in the binding site, we are interested in analysing the interaction of the new series with some EGFR mutants, such as the famous T790M mutation [26,27]. This mutation, among others, is responsible for the development of resistance against Lapatinib as well as other Tyrosine Kinase Inhibitor's (TKIs) [26,27]. Previous studies have shown that acquired Lapatinib resistance due to the T790M (gate keeper) mutation is mostly driven by either severe steric clashes or disruption of the H-bonding network due to the surrounding water molecules [28-30]. Moreover, the T790M mutation was linked to increased affinity to ATP and only those inhibitors having a covalent binding mode can still be effective [31].

In the two previous studies, we have identified the inhibitor M19 and M20 to be the most active against EGFR and HER2 [10,11]. Herein, we are giving the results of a primary *in silico* screening of a designed compound library of 200 inhibitors. We will also represent one of the most promising members of the *in silico* screened compound library, which is the T9 inhibitor. Additionally, we will include SYR and Lapatinib as references for the simulation. We will also give some primary results for MD simulations of Lapatinib, M19 and T9 against the T790M EGFR mutant. In total, 9 complexes are subjected to long MD simulation; i. Lapatinib, M19 and T9 with wild type EGFR (WT), ii. Lapatinib, M19 and T9 with the T790M mutant EGFR(MUT) and iii. SYR, M19 and T9 with HER2 (WT). For simplicity, the wild type EGFR will be denoted as EGFR(WT) and the mutant strain will be denoted as EGFR(MUT). Wild type HER2 will be denoted as HER2. In addition to MD simulations, we have carried out binding free energy calculations based either on a single snapshot (Prime-MM/GBSA) or ensemble of snapshots (AMBER-MM/PB(GB)SA). To account for the protein polarization effect, we have also rescored the relaxed MD complexes using QM/MM rescoring.

## 2. METHODS AND COMPUTATIONAL DETAILS

### 2.1. Design Rationale

Figure 1 represents the 2D structure of Lapatinib, a dually acting anti-EGFR/HER2 drug, SYR which is an HER2 inhibitor, in addition to the two leads that have been identified previously M19 and M20. We also include the 2D structure of the most active inhibitors of the new designed library identified *in silico* (T9). The *in vitro* measured $IC_{50}$ of M19 is (1.935 μM: EGFR and 1.035 μM: HER2). The new designed inhibitors will benefit from the main skeleton of M19 in addition to substitution at four different sites where we believe that substitution may give rise to more effective inhibitors. At this stage, we decided not to go to severe modifications on the main nucleus itself to keep essential binding motifs.

As we have identified previously, the binding of the new series to either EGFR or HER2 is very similar, the N1 atom of the quinazoline ring binds to Met793 (EGFR) or Met801 (HER2). The bulky 4-anilinoquinazoline side group binds to the side pocket formed by displacement of the Cα-helix. The two terminal tails at positions $R_1$ and $R_2$ push the quinazoline ring for better filling of the hinge region and also form lipophilic interactions with the nearby residues [10,11].

Keeping in mind that the lipophilic vdW interaction terms is the most important term in these class of receptors, the primary focus on choosing different substituents is the halogen atoms at position $R_3$ and $R_4$ (-F, -Cl,-Br, -I, -CF3). For $R_1$ and $R_2$ positions, we increase the chain length from one carbon (-methoxy) to three carbons (-propoxy). We also utilize the erlotinib and Lapatinib side chain tails instead of the pure alkoxy group substituents. The correctly docked inhibitors are given in supplementary data. Note that this list contains also modifications for the second most active anti-EGFR ligand identified previously which is M20.

### 2.2. Docking and scoring calculations

For more details about the docking preparation and setup, please refer to our previous work [11]. All ligands are docked against the EGFR crystal structure (PDB code: 1xkk) using the same settings and ranked according to the XP_LipophilicEvdW term. Only inhibitors that satisfy the presence of a two predetermined H-bonds constraints (with Met793 and the conserved water molecule) are kept. The XP_LipophilicEvdW term was able to give the highest correlation with the experimentally measured inhibitory activity as we have shown previously ($r_{(Pearson\ correlation)}$=0.78 and $r^2$= 0.61) [10,11]. The designed T9 inhibitor has the highest XP_LipophilicEvdW score and is chosen for more detailed computational investigation. Our primary results suggest that T9 is able to form a more stable interaction with both EGFR and HER2. T9 is also much more able than M19 to retain the interactions with the conserved water molecules observed in several crystal structures of kinases complexed with their corresponding inhibitors [32-35]. The T9-HER2 complex is obtained by mutating the M19 inhibitor in the M19-HER2 complex obtained previously using the IFD protocol [10]. This is followed by a quick relaxation with Macromodel [36] to remove steric clashes.

For those complexes that are subjected to the MD simulations, physics based rescoring is carried out applying both MM and QM/MM techniques. For MM rescoring, we apply a single, end point rescoring for the final trajectory snapshot produced from the MD simulations and using the Prime/MM-GBSA in the Schrodinger suite [37-39]. The complexes are also rescored applying the much more sophisticated AMBER/GB(PB)SA module in AMBER12 [40,41]. The AMBER/GB(PB)SA module takes the advantage of statistical averaging over many potential configurations. Every second frame of the produced 8000 frames has been chosen, i.e., a total of 4000 simulation snapshots have been utilized. For the QM/MM rescoring and because of the prohibitive computational expense, in DFT/MM rescoring we use only the ligand as the QM subsystem as implemented in the QM/MM-PBSA script of Schrodinger and solvation effect is ignored in that case. The QM level is chosen to be the B3LYP/LACVP* level and the OPLS2005 force field is used to describe the MM subsystem. For a more realistic QM/MM binding energy estimation, the AMBER QM/MM-GBSA calculations are utilized and the QM subsystem is extended to the ligand in addition to the surrounding residues within 5Å from the ligand. The dispersion corrected AM1 (AM1-D) [42,43] Hamiltonian is utilized for the QM subsystem and for the MM subsystem, the ff03 force field is used [44].

### 2.3. Molecular dynamic setup

Nine inhibitor-protein complexes (see above) obtained from the previous rigid (EGFR) and IFD (HER2) runs are subjected to extensive molecular dynamic simulations. For the Lapatinib-EGFR (PDB code: 1XKK) and the SYR-HER2 (PDB code: 3PP0) complexes, the original crystal structures are used [45,46]. The structure preparation and the following MD simulations are performed using AMBER12 software package [41] applying the ff03 force field [44,47]. Single point calculations of the corresponding inhibitors are performed at the B3LYP/cc-pVTZ level of theory in ether ($\varepsilon$=4.2) solvent applying the IEF-PCM model [48] using the Gaussian 09 program [49]. The inhibitor charges and other parameters are obtained using the RESP fitting [50] procedures and the general AMBER force field (GAFF) [51]. The complexes are then solvated with a box of TIP3P [52] water with a buffer size of 15Å and are neutralized by counter ions.

Each system is then subjected to four consecutive minimization steps. In each step, water molecules and ions are allowed to move freely for a 1000 steps of steepest descent minimization followed by 4000 steps of conjugate gradient minimization. During minimization, protein and inhibitor atoms are constrained to their original positions by a force constant of 100 kcal·mol$^{-1}$Å$^{-2}$, then gradually releasing the force constraints to 50, 5 and zero (no constraints) kcal·mol$^{-1}$Å$^{-2}$, respectively. Following minimization, two consecutive steps of heating and equilibration are performed. First; each system is gradually heated in the NVT ensemble from 0°K to 300°K for 30 ps with a time step of 1 fs, applying a force constant of 10 kcal·mol$^{-1}$Å$^{-2}$ on the protein and inhibitor coordinates, and using Langevin dynamics with the collision frequency $\gamma$ of 1 $ps^{-1}$ for temperature control. A further 1 ns simulation in the NPT ensemble is performed to equilibrate the system density applying a time step of 2 fs, which requires the use of SHAKE algorithm [53] to constrain all bonds involving hydrogen atoms. The temperature is controlled using Langevin dynamics with the collision frequency $\gamma$ of 1 $ps^{-1}$ and is kept at 300°K. The pressure is kept at 1 bar applying a Berendsen barostate with a pressure relaxation time of 1 ps. Each system is again relaxed in the NVT ensemble for another 2ns, in the first 1ns, heavy atoms are constrained by a force constant of 10 $kcal \cdot mol^{-1}Å^{-2}$. Each system is then subjected to a 10 ns "warming up" simulation, followed by 4 production simulations of 20 ns long each i.e., 80(4*20) ns in total applying the NVT ensemble at 300°K and using Berendsen temperature control [54]. In all simulation steps, long-range electrostatics are computed using the particle mesh Ewald (PME) [55] with a non-bonded cut-off of 12 Å. The edge effect is removed by applying periodic boundary conditions. All MD simulations are carried out using the PMEMD module of AMBER12 [41]. Binding energies are calculated from the MD trajectories using the PBSA module in AMBER12.

## 3. RESULTS AND DISCUSSION

Almost all TK inhibitors are either competitive or non-competitive inhibitors for the evolutionary conserved ATP binding site of the Tyrosine Kinase systems, EGFR and HER2. This ATP binding site is located in a deep cleft connecting the N-lobe and C-lobe and is conventionally known as "Hinge region". Owing to their high flexibility, available X-ray crystal structures of kinases complexed with their inhibitors show high degree of inhibitors diversity with respect to the molecular structure of the inhibitors. This high flexibility also imposes challenges to the molecular modelling process giving more room for the modelling scientist experience. There are several methods for TKIs classifications [56-58]. Briefly speaking, TKIs are classified into two main classes. The first class of inhibitors are those targeting either the active (DFG-in) or the inactive (DFG-out) state of the protein [45,59]. The second class of inhibitors are those bind to the (DFG-in) state in addition to a selectivity pocket formed as a result of large shift of the Cα-helix which also represents an inactive state of the protein [58]. This large shift creates a deep pocket in the back of the protein giving more room to accommodate bulky groups in the corresponding inhibitors.

Other inhibitors can also bind the ATP pocket and nearby allosteric sites [58]. An example from the first class is Gefitinib (Iressa®) which is able to bind the active (DFG-in) state of the protein [60,61]. This inhibitor, among others, is characterized by the presence of small side group at the 4-anilino site. An example of the second class is Lapatinib (Tykerb®) [62,63]. In the contrary of Gefitinib, Lapatinib possess a bulkier group at the 4-anilino site. This bulky group is accommodated by a deep pocket created as a result of a large shift of the Cα-helix [45,59]. The X-ray crystal structure of HER2 complexed with the inhibitors (SYR) has been released recently [64]. Previous modelling studies for HER2 have utilized homology modelling and using EGFR as a template owing to its high sequence similarity to HER2 (81%) [29,30].

In general, inhibitor binding to either EGFR or HER2 is driven by strong lipophilic vdW forces in addition to strong H-bond with a conserved Met residue at the hinge region with the N1 atom of the quinazoline ring. Other studies have also identified the importance of the conserved water molecule typically seen in many crystal structures of kinases co-crystalized with their corresponding ligands. However, it's an experimental fact that the binding energy contribution for this water molecule is essentially small and some of the active inhibitors previously reported lack the N atom at the 3 position (N3) that is responsible for the interaction with this water molecule [65,66]. In certain cases, an extra cyano group at this site reduces the need of this N3 atom via direct interaction with the nearby Thr [67,68]. For the new series under investigation, our primary results suggest that substitution of ring "**C**" by a halogen is essential to obtain such conserved water H-bond interaction, and this may be because of steric reasons.

### 3.1. Docking and scoring

Table 1 reports the XP_Gscore and the XP_LipophilicEvdW docking scores of the 10 most active *in silico* designed inhibitors together with the substitution groups at $R_1$, $R_2$, $R_3$ and $R_4$. Previously, we have shown that the lead structure M19 was identified to be the top active inhibitor against both EGFR and HER2 [10,11]. For this section, we consider only EGFR as a result of data transferability which originates from high sequence similarity (81%). Our discussion is limited to the M19 modifications as M19 is the dually acting inhibitor whereas M20 shows a very weak anti-HER2 inhibitory effect ($IC_{50}$ against EGFR: 2.582 μM and against HER2 99.96 μM) [10,11]. Given the exceptional power of the XP_LipophilicEvdW score to qualitatively identify the top active anti-EGFR as we have shown previously, we will mainly consider that score in the present docking study. The derivatives also appear in the table according to this score.

As we can see in Table 1, all modifications indeed showed a higher XP_LipophilicEvdW score. Interestingly, pure alkoxy substituents at $R_1$ and $R_2$ positions are more favoured than hetero substituted alkyl group, such as the ethoxy-methoxy group present in Erlotinib (as in T8e with XP_LipophilicEvdW score of -9.28 and T11e with XP_LipophilicEvdW score of -9.20) or the substituted Furyl group present in Lapatinib (as in T12l with an XP_LipophilicEvdW score of -8.96 ). We didn't attempt to test electron withdrawing groups (EWG) at these two sites as previous studies have shown that EWG at these two sites of the anilinoquinazoline scaffold tend to decrease the anti-tumour activity significantly [69,70]. Also, it seems that optimal activity is achieved with Chlorination at $R_3$ and $R_4$ positions as the case in T9 (XP_LipophilicEvdW score of -9.40). Again, synthesis and biological investigation for this new series is warranted to build such solid conclusions.

## 3.2. Molecular dynamic simulations

### 3.2.1. General binding remarks

Figure 2 represents the root mean square deviations plots (RMSDs) for the protein Cα atoms for the nine complexes under study through the 80 ns production simulations. As we can see in the figures, the inhibitors-protein complexes show an overall reasonable RMSD values and slight discrepancies are expected as a result of the known flexibility of kinases. As mentioned before, Lapatinib binds to the ATP binding site of either EGFR and/or HER2 receptors. The binding is driven mainly by vdW lipophilic interactions as have been shown before. An essential requirement for binding is the presence of a nitrogen atom at position no. 1 of the heterocyclic ring (N1). This heterocyclic ring may be a quinoline, quinazoline, pyridopyrimidine or even a mono-cyclic ring [71]. This nitrogen binds strongly to a Met residue at the hinge region (Met793 in EGFR and Met801 in HER2). Derivatives without N atom at this position are almost inactive [71,72]. Nitrogen substitution at position 3 (N3) guarantees the binding to a water molecule seen frequently in kinase crystal structures. This water molecule forms a bridge between the N3 of the inhibitor and nearby residues side chains, such as Cys and Thr. It seems also that this water molecule, together with other water molecules, forms a network of H-bonds around the inhibitor, this site is referred to often as the S1 site. Although such water molecule is not essential for the anti-EGFR and/or anti-HER2 activity, recent studies link some binding problems against kinases to the disruption of such favouring interaction [12,30]. Another two potential sites for the presence of water molecules are the S2 site in which water molecule bind to the aniline hydrogen, the third site S3 is linked to the nearby Gln amino acid. Figure 3 shows the potential locations of these three sites around the inhibitors under study.

### 3.2.2. Water molecules occupancies

Subsequent analysis of the MD trajectories shows that the H-bonds occupancies of these three sites are different for different inhibitors-protein complexes. Table 2 represents the % occupancy of the potential water mediated H-bonds as a function of the simulation time. In all cases, an upper distance cut-off of 3.5 Å and a lower angle cut-off of 120° are used. Figure 3 displays potential water occupancy sites (S1, S2 and S3) among the studied inhibitors. As there's a possibility for the two hydrogens of the same water molecule or more than one water molecule to interact simultaneously, one may expect that the occupancy may be even more than 100%.

In all situations, the most conserved occupancy is observed with the S3 site. The T9-HER2 complex achieves the highest value of 123.54% of the simulation time for this site (S3). We should stress here that neither of these water molecules have been got from the original crystal structure nor constrained to their 3D coordinates. However, as the occupant water molecules at this site (S3) don't directly interact with the inhibitor, this site may have the least influence among the three observed water sites. For S2 site, %occupancy values for all complexes are comparable with the exception of the SYR-HER2(WT) complex of having Zero% occupancy at this site possibly because of the potential formation of an intramolecular H-bond.

The most interesting water %occupancy is observed with the S1 site. The highest occupancies are achieved with either SYR-HER2(WT) with a %occupancy of 97.54 followed by Lapa-EGFR(WT) with a value of 94.50. For the new inhibitors, both the lead (M19) and the optimized structure (T9) show an interesting behaviour. M19 seems to completely lack such interaction in all situations with a maximum value of 1.71% for its complex with EGFR(WT). Such low occupancy value can be considered essentially insignificant. In contrast to M19, T9 seems to have a much higher occupancy at the S1 site of about 91.05% with HER2(WT) and a value of 70.03% with EGFR(WT). As expected and as a result of the increased hydrophobicity around the S1 site due to the T790M mutation, water occupancy at this site is significantly reduced from 94.50% in Lapa-EGFR(WT) to 28.69% in Lapa-EGFR(MUT) and from 70.03% in T9-EGFR(WT) to Zero% in T9-EGFR(MUT). A quick look on the literature show that "almost" all effective anilinoquinazoilne EGFR and/or HER2 competitive inhibitors, particularly the currently marketed inhibitors, are substituted at ring "**C**" mostly with an electron withdrawing groups, such as halogen and/or ethyne groups [66,69,73]. In the new series, the lead structure M19 doesn't have but the modified one T9 does have such substitution, so, T9 possess such water mediated interaction at S1 as we showed before. Our primary investigation showed that this may be because of steric reasons. Because of the existence of different rotomeric conformations of the side chains of nearby residues responsible for binding to this bridging water molecule at the S1 site, i.e. Thr790 and Thr854, such interaction is possible in Lapatinib and T9 but not possible in M19. The same is true for inhibitors-HER2 complexes and the differences between rotomeric structures of Thr798 and Thr862 are responsible for the existence of high water occupancy at the S1 site for SYR (97.54%) and T9 (91.05%) but as low as 0.11% for M19.

### 3.2.3. Binding to the conventional hinge residues

As mentioned before, inhibitor binding to both EGFR and HER2 is driven mainly by lipophilic vdW interactions. The strength of this binding can be *in silico* assessed by calculating the energy of binding as well as the stability of the formed complexes. For the inhibitors under study, all are able to form a stable interaction with the conserved hinge Met residue (Met793 for EGFR and Met801 for HER2). Previous studies have shown that for EGFR receptors, acquired resistance due to the T790M mutation prevents stable interaction with competitive inhibitors, including Lapatinib. This has been previously proposed to be attributed to several reasons; (i) steric clashes, (ii) increased sensitivity to ATP and (iii) disruption of the surrounding waters mediated H-bonds [29,30]. Indeed, we show that the surrounding water mediated H-bonds network is severely disrupted in case of the T790M mutation. However, we believe that such disruption is not enough alone to explain the huge reduction in the ability of Lapatinib to inhibit EGFR as a consequence of this mutation. The anti-tumour activity of Lapatinib against EGFR is severely decreased going from the EGFR(WT) with an $IC_{50}$ of 0.022 μM to EGFR(MUT) with an $IC_{50}$ of 3.3 μM [29]. We believe that a combination of the aforementioned explanations is the most likely reason.

Figure 4 shows the fluctuation of the N1-Met793 (EGFR) or the N1-Met801 (HER2) during the 80 ns simulation time. This H-bond is the most important anchor and, as we mentioned before, inhibitors lack such H-bond are inactive. Indeed, in all cases the fluctuation is minimal and this H-bond can be considered very stable. An exception is the Lapatinib complex with the EGFR(MUT) that shows a large fluctuation after about 60 ns to about 5 Å. This may shed some light on the possible instability of Lapatinib-EGFR(MUT) complex. However, ligand escape during the conventional simulation time for EGFR(MUT) is not expected giving the limitations of the current modelling tools and the complicated energy landscape [74]. This has been previously discussed for the EGFR T790M mutant binding to Gefitinib [74]. In all other complexes, the average value of this H-bond ranges between 2-2.5 Å.

### 3.2.4. Binding to the additional hinge residues

Previously we have demonstrated the ability of the new inhibitors to bind to additional hinge residues in both EGFR and HER2 binding sites, and due to the high sequence similarity between both receptors, the locations of such interactions are similar [10,11]. In general, the most important interactions are the two charge assisted H-bonds formed with Asp855 and Lys745 (EGFR) or Asp863 and Lys753 (HER2) and mediated by the N-H group of ring "D" and the extra imino group on the same ring, respectively. Because of structural reasons, neither SYR nor Lapatinib are able to form such charge assisted interactions. However, another possibility for SYR interaction with Asp863 is via the terminal hydroxy group at the other side of the molecule. This interaction is indeed observed in the crystal structure itself and is formed during the simulation. Figure 5 displays the interaction of T9 and M19 with the three receptor types (EGFR(WT), EGFR(MUT) and HER2). As such static representation may be misleading as it doesn't provide any dynamic information, Table 3 reports the %occupancies of such interaction as a function of the simulation time. As we can see in the table, the most stable interactions are the T9-Asp863(HER2) and the M19-Lys753(HER2) interactions with an occupancy of 71.03% and 66.86%, respectively. T9 seems to keep a higher %occupancy for these two interactions with EGFR(MUT) than its corresponding interaction with EGFR(WT). Also, the occupancy of the M19 interaction with Lys745-EGFR(MUT) is given by 66.2% whereas its interaction with Lys745-EGFR(WT) is given by 41.96%. In contrast to the M19 interaction with Lys745 which is higher in the EGFR(MUT) than the EGFR(WT), the occupancy of M19 H-bond interaction with Asp855 is higher in the EGFR(WT) (12.51%) than with EGFR(MUT) (1.78%). Occasionally, M19 is able to form an H-bond with the amide oxygen of Asp863 in HER2.

This demonstrated ability of the new inhibitors to bind to these extra residues is a good point that may favour their future development. It's known that some of the very challenging mutations that result in resistance to current EGFR/HER2 are those present in the vicinity of the hinge region. An example of such mutations is the stubborn T790M mutation which is known to be present in more than 50% of all patients whose tumours showed an acquired resistance for TKIs [75,76]. Formation of very strong interactions with additional residues may retain the antitumor activity even in the EGFR/HER2 resistant strains including the T790M mutation. Although it has been shown that irreversible inhibitors may be the agents of choice for the resistant strains, toxicity concerns may be a problem for these agents [28,77,78].

## 3.3. Binding energy calculations

### 3.3.1. Total binding energy

Table 4 reports the binding energy scores for the compounds under study at different levels of QM/MM and pure-MM levels of theory. In general, MM based methods perform well in comparison with the QM/MM based methods. With the exception of the pure gas phase $\Delta E_{QM/MM}$ binding energy scores, T9 exhibits higher binding energy scores than M19 in all situations. We believe that this rather questionably high $\Delta E_{QM/MM}$ binding energy score for M19 (-141.54 kcal·mol$^{-1}$Å$^{-2}$) is a direct consequence for the absence of the solvent screening effect and the statistical inaccuracies as well. The absence of such screening effect amplifies certain intermolecular interactions significantly. However, further investigation is required to identify the origin of such exaggerated score. The introduction of the conserved water molecule seems to show some increase to the overall binding score and particularly to the $\Delta E_{col}$ term as we will discuss in the following section. Direct comparison between binding energy scores for ligands having different scaffolds should be conducted with caution as a result of different entropic terms and dissociation rates.

Overall, Lapatinib shows a higher binding energy scores for the mutant (MUT) rather than the wild-type (WT) strain of EGFR. Admittedly enough, only the dispersion corrected AM1 Hamiltonian (AM1-D) predicts M19 to be a better binder for HER2 (-162.85 kcal·mol$^{-1}$) than EGFR (-133.60 kcal·mol$^{-1}$). This is consistent with experimentally measured $IC_{50}$ for M19 against HER2 (1.035 μM) and EGFR (1.935 μM) [10,11]. Also, we tried to include all surrounding residues within the binding site in the QM region as no much QM accuracy is expected to be gained considering only the ligand as the QM subsystem. Other binding energy scores show discrepancies as a result of different modelling techniques and theory levels.

### 3.3.2. Component contribution to total binding energy

Table 5 reports the individual binding energy components calculated using the AMBER/MM-PBSA tool. The total binding energy mentioned in that table ($\Delta E_{tot}$) doesn't include the desolvation penalties included in Table 4, i.e., we can consider that $\Delta E_{tot}$ is a "gas phase" binding energy. As we can see in the table and in all complexes, the $\Delta E_{vdW}$ term is the major component. For the EGFR(WT) complexes, T9 shows enhanced both $\Delta E_{vdW}$ and $\Delta E_{col}$ binding energy components (-83.09 kcal·mol$^{-1}$ and -31.57 kcal·mol$^{-1}$, respectively) with respect to that for M19 (-75.99 kcal·mol$^{-1}$ and -24.20 kcal·mol$^{-1}$, respectively) and Lapatinib (-76.42 kcal·mol$^{-1}$ and -29.16 kcal·mol$^{-1}$, respectively). Interestingly, including the conserved water molecule (S1) through the 1WAT protocols enhances the $\Delta E_{col}$ term, with little or no effect on the $\Delta E_{vdW}$ term. This results in an increase in the $\Delta E_{col}$ term by 3.49 kcal·mol$^{-1}$ and 4.22 kcal·mol$^{-1}$ for both T9 and Lapatinib, respectively.

For the EGFR(MUT) complexes, we notice an increase in both the $\Delta E_{vdW}$ and $\Delta E_{col}$ terms for Lpatininib (-85.88 kcal·mol$^{-1}$ and -39.15 kcal·mol$^{-1}$, respectively) with respect to that for T9 (-81.23 kcal·mol$^{-1}$ and -26.46 kcal·mol$^{-1}$, respectively) and M19 (-73.27 kcal·mol$^{-1}$ and -22.27 kcal·mol$^{-1}$, respectively). The Lapatinib binding energy components in the EGFR(MUT) seem to be even higher than their corresponding values for the EGFR(WT). Such behaviour has been pointed out previously by Huang et al. [30].

For the HER2 complexes, T9 seems to have the highest contribution from the $\Delta E_{vdW}$ term which is given by -76.19 kcal·mol$^{-1}$ compared to M19 (-70.56 kcal·mol$^{-1}$) or SYR(-68.27 kcal·mol$^{-1}$). The $\Delta E_{col}$ term for SYR seems to be greatly enhanced by the introduction of the water molecule ($\Delta\Delta E_{col}$ = 5.4 kcal·mol$^{-1}$), in contrast to that gain for T9 ($\Delta\Delta E_{col}$ = 3.4 kcal·mol$^{-1}$).

### 3.3.3. Per-residue binding energy decomposition

Individual binding energy components on per-residue decomposition analysis are also shown in Figure 6. We choose only the most important residues to show. Choice is based primarily on the residue contribution to the $\Delta E_{vdW}$ term for the new inhibitors, T9 and M19. Figure labels represent the exact binding contribution of each selected amino acid to the corresponding term. As we can see in the figure, both Asp855(863) and Lys745(753) have a high contribution to the binding energy for both the $\Delta E_{vdW}$ and $\Delta E_{col}$ terms. Also, in all complexes, T9 got the highest contribution from Met793(801) which can be considered the most important anchoring point.

For the EGFR(WT) complexes, the Asp855 contribution for the $\Delta E_{vdW}$ term is higher for M19 (-3.48 kcal·mol$^{-1}$) than for T9 (-3.35 kcal·mol$^{-1}$), on the other hand, the Lys745 contribution is higher for T9(-2.28 kcal·mol$^{-1}$) than for M19 (-1.27 kcal·mol$^{-1}$). Discrepancies between T9 and M19 for the $\Delta E_{col}$ term regarding these two amino acid residues seem to be non-specific in nature as they are not directly correlated with the H-bond occupancies (Table 3). Other possibility may be because of the nature of the involved H-bond itself. What makes the first assumption more acceptable is the slightly high $\Delta E_{col}$ contribution of Asp855 residues to Lapatinib binding although Lapatinib doesn't form a direct specific H-bond interaction with this residue. The Asp855 contribution to the $\Delta E_{col}$ term for Lapatinib binding to the EGFR(MUT) is given by -7.16 kcal·mol$^{-1}$, again there's no direct specific H-bond interaction between Lapatinib and Asp855. Also, for the EGFR(MUT), T9 and M19 possess a higher contribution from Asp855 to the $\Delta E_{vdW}$ term (-2.52 kcal·mol$^{-1}$and -3.63 kcal·mol$^{-1}$, respectively) than Lapatinib (-2.1 kcal·mol$^{-1}$).

For HER2 complexes, it seems that M19 get the highest contribution for the $\Delta E_{vdW}$ from the Phe864 residue (-3.06 kcal·mol$^{-1}$). This residue is particularly interesting as neither SYR nor Lapatinib (it's the Phe856 in EGFR) show a significant contribution to the overall $\Delta E_{vdW}$ term from this residue. Indeed, the good contribution for this residue to the $\Delta E_{vdW}$ term comes as a consequence of the additional aromatic ring in both T9 and M19. Future development to this series will include this residue as an important target. For the $\Delta E_{vdW}$ term, it seems that M19 gets a higher contribution from the Asp863 residue (-2.35 kcal·mol$^{-1}$) than T9 (-1.97 kcal·mol$^{-1}$), however, T9 got a higher contribution from Lys753 (-2.03 kcal·mol$^{-1}$) than M19(-0.94 kcal·mol$^{-1}$) for the same term. For the $\Delta E_{col}$ term, M19 shows higher contributions from both Asp863 (-4.45 kcal·mol$^{-1}$) and Lys753 (-5.94 kcal·mol$^{-1}$) than T9 (-3.03 kcal·mol$^{-1}$ and -5.2 kcal·mol$^{-1}$ , respectively).

Now this question arises, do we expect the *in silico* designed dually acting inhibitor (T9) to be of a better inhibitory profile than Lapatinib. This question is not easy to answer given the limitations of the currently available modelling tools. Assessment of the ligand-protein binding requires more than just the binding energy scores. For example, previous studies have shown that although Gefitinib may possess a higher *in silico* binding energy than lapatinib against EGFR, the Lapatinib-EGFR complex is much more stable [58]. Lapatinib forms an "almost" irreversible complex with EGFR with a $t_{1/2}$ of about 224 minutes [29]. Further development, synthesis and biological testing are warranted in order to achieve the best outcome.

## 4. CONCLUSIONS

In this study, we have characterized the binding modes and exact mechanisms of inhibition of a recently synthesized 4-anilinoquinazoline series against both wild type and T790 mutant EGFR in addition to wild type HER2. It seems that substitution at ring "**C**" is essential to achieve optimum interaction with the conserved water molecule observed frequently in kinase crystal structures. The new series also demonstrated an ability to bind additional residues in the binding site utilizing direct H-bonding interactions, namely; Asp855 and Lys745 for EGFR and Asp863 and Lys753 for HER2. Inclusion of the binding site water in calculating the total binding energy resulted in an increase of about 3-5 $kcal \cdot mol^{-1}$ and this enhancement was mainly due to increase in the $\Delta E_{col}$ term. Both MM and QM/MM based rescoring methods perform well in predicting the overall trends in binding energy.

## ACKNOWLEDGEMENTS

MA acknowledges the Swinburne University Postgraduate Research Award (SUPRA). We thank the Victorian Partnership for Advanced Computing (VPAC) and Swinburne University supercomputing (Green/Gstar) for the support on the computing facilities. The National Computational Infrastructure (NCI) at the Australian National University and the Victorian Life Sciences Computation Initiative (VLSCI) on its Peak Computing Facility at the University of Melbourne (an initiative of the Victorian Government, Australia) under the Merit Allocation Scheme (MAS) are acknowledged.

**Table 1**: R group modifications of the *in silico* designed compound library with the docking scores.

| Title | Substituents | | | | XP GScore | XP_LipophilicEvdW | %inhibition [a] | $IC_{50}$ (μM) |
|---|---|---|---|---|---|---|---|---|
| | $R_1$ | $R_2$ | $R_3$ | $R_4$ | | | | |
| T9 | $-(CH_2)_2-CH_3$ | $-(CH_2)_2-CH_3$ | -Cl | -Cl | -9.02 | -9.40 | | |
| T16 | $-(CH_2)_2-CH_3$ | $-(CH_2)_2-CH_3$ | -Cl | $-CF_3$ | -8.88 | -9.39 | | |
| T11 | $-(CH_2)_2-CH_3$ | $-(CH_2)_2-CH_3$ | -Br | -Br | -8.89 | -9.38 | | |
| T1 | $-(CH_2)_2-CH_3$ | $-(CH_2)_2-CH_3$ | -Cl | -Br | -8.92 | -9.33 | | |
| T8e | $-(CH_2)_2-O-CH_3$ | $-(CH_2)_2-O-CH_3$ | -Br | -Cl | -8.62 | -9.28 | | |
| T11e | $-(CH_2)_2-O-CH_3$ | $-(CH_2)_2-O-CH_3$ | -Br | -Br | -8.74 | -9.20 | | |
| T12 | $-(CH_2)_2-CH_3$ | $-(CH_2)_2-CH_3$ | -F | -Br | -10.05 | -9.20 | | |
| T3 | $-(CH_2)_2-CH_3$ | $-(CH_2)_2-CH_3$ | -Cl | -F | -9.03 | -9.18 | | |
| T8 | $-(CH_2)_2-CH_3$ | $-(CH_2)_2-CH_3$ | -Br | -Cl | -11.33 | -8.99 | | |
| T12l | 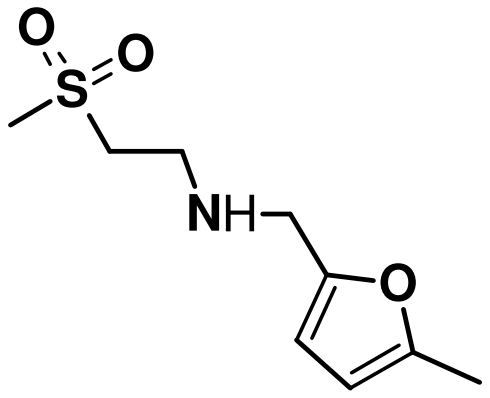 | -H | -F | -Br | -7.76 | -8.96 | | |
| M19 | $-CH_3$ | $-CH_3$ | -H | -Br | -10.78 | -8.40 | 72 | 1.935 |

[a] At 5 μM concentration.

**Table 2**: %Hydrogen-bond occupancy analysis of the three observed water sites as a function of the simulation time

| Inhibitor-EGFR(WT) | | | Inhibitor-EGFR(MUT) | | | Inhibitor-HER2(WT) | | |
|---|---|---|---|---|---|---|---|---|
| T9 | S1 | 70.03 | T9 | S1 | Zero | T9 | S1 | 91.05 |
| | S2 | 85.54 | | S2 | 70.09 | | S2 | 67.44 |
| | S3 | 89.89 | | S3 | 91.80 | | S3 | 123.54 |
| M19 | S1 | 1.71 | M19 | S1 | 1.10 | M19 | S1 | 0.11 |
| | S2 | 41.61 | | S2 | 69.25 | | S2 | 58.20 |
| | S3 | 61.09 | | S3 | 78.14 | | S3 | 111.48 |
| Lapa | S1 | 94.50 | Lapa | S1 | 28.69 | SYR | S1 | 97.54 |
| | S2 | 94.77 | | S2 | 60.84 | | S2 | Zero |
| | S3 | 114.90 | | S3 | 69.14 | | S3 | 113.65 |

**Table 3**: %Hydrogen-bond occupancy analysis of the additional potential hinge residues interaction.

| Inhibitor-EGFR(WT) | | | Inhibitor-EGFR(MUT) | | | Inhibitor-HER2(WT) | | |
|---|---|---|---|---|---|---|---|---|
| T9 | Asp855 | 14.49 | T9 | Asp855 | 66.51 | T9 | Asp863 | 71.03 |
| | Lys745 | 13.02 | | Lys745 | 31.84 | | Lys753 | 56.92 |
| M19 | Asp855 | 12.51 | M19 | Asp855 | 1.78 | M19 | Asp863 | Zero |
| | Lys745 | 41.96 | | Lys745 | 66.2 | | Lys753 | 66.86 |

**Table 4**: Experimental %inhibition together with simulated binding energies and scores for the MD produced complexes (kcal·mol$^{-1}$).

| Title | $\Delta E_{QM/MM}$ [a] | AMBER-QM-MM/GBSA [b] | Prime-MM/GBSA [c] | AMBER-MM/GBSA [d] | AMBER-MM/PBSA [d] | $IC_{50}$ (μM)[e] | %inhibition[f] |
|---|---|---|---|---|---|---|---|
| **EGFR(WT)** | | | | | | | |
| M19 | -141.54 | -133.60 | -136.14 | -65.63 | -95.29 | 1.935 | 72 |
| T9 | -97.71 | -139.86 | -152.83 | -69.95 (-73.58) | -103.44 (-108.46) | | |
| Lapatinib | -116.86 | -91.16 | -128.37 | -68.12 (-71.74) | -99.93 (-103.57) | 0.022 | |
| **EGFR(MUT)** | | | | | | | |
| M19 | -84.25 | -121.18 | -130.75 | -61.83 | -92.18 | | |
| T9 | -115.27 | -139.06 | -153.56 | -70.45 | -102.07 | | |
| Lapatinib | -118.44 | -96.38 | -142.36 | -75.16 | -105.62 | >3.3[g] | |
| **HER2** | | | | | | | |
| M19 | -86.33 | -162.85 | -116.14 | -59.34 | -93.34 | 1.035 | 85 |
| T9 | -104.36 | -164.23 | -159.58 | -66.28 (-69.04) | -100.38 (-104.36) | | |
| SYR | -107.75 | -77.73 | -136.64 | -62.38 (-66.31) | -91.36 (-96.02) | 0.011 | |

[a] Calculations are carried out using the QM/MM-PBSA script from Schrödinger and ignoring solvation, the B3LYP/LACVP* //OPLS2005 level of theory is applied, QM substructure composed of inhibitor atoms only.
[b] QM system composed of ligand and the nearby residues within 5 Å, QM system was treated at the dispersion corrected AM1 Hamiltonian (AM1-D) and MM system was treated at the AMBER-ff03 force field.
[c] Residues within 6 Å of the inhibitor are treated flexibly.
[d] Values between brackets are those obtained upon inclusion of the conserved water molecule in the calculation.
[e] See Ref. [10,11].
[f] At 5 μM concentration, See Ref. [10,11].
[g] See Ref. [29].

**Table 5**: Decomposed binding energies of the MD studied complexes (kcal·mol$^{-1}$).

| Title | **$\Delta E_{vdW}$** [a] | **$\Delta E_{col}$** [a] | **$\Delta E_{tot(vdW+col)}$** [a] |
|---|---|---|---|
| **EGFR(WT)** | | | |
| M19 | -75.99 | -24.20 | -100.19 |
| T9 | -83.09(-84.07) | -31.57(-35.06) | -114.66(-119.13) |
| Lapatinib | -76.42(-77.15) | -29.16(-33.38) | -105.58(-110.53) |
| **EGFR(MUT)** | | | |
| M19 | -73.27 | -22.27 | -95.54 |
| T9 | -81.23 | -26.46 | -107.69 |
| Lapatinib | -85.88 | -39.15 | -125.03 |
| **HER2** | | | |
| M19 | -70.56 | -26.75 | -97.31 |
| T9 | -76.19(-76.96) | -20.23(-23.63) | -96.42(-100.59) |
| SYR | -68.27(-68.69) | -24.59(-29.99) | -92.86(-98.68) |

[a] Values between brackets are those obtained upon inclusion of the conserved water molecule in the calculation.

## Figure captions

**Figure 1**: The 2D chemical structures of the compounds under study.

**Figure 2**: Cα backbone atoms RMSDs plots for the studied complex during the 80 ns MD simulations.

**Figure 3**: Potential binding poses and water orientations of the studied complexes.

**Figure 4**: Hydrogen bond distance profiles between the inhibitors and the Met hinge residue (Met793 in EGFR and Met801 in HER2) during 80 ns MD simulation.

**Figure 5**: 3D interaction with the conserved hinge region Met residue and the additional residues Asp and Lys.

**Figure 6**: Per-residue binding energy decomposition of the 9 MD runs using the MMPBSA module of AMBER.

**Figure 1**: The 2D chemical structures of the compounds under study.

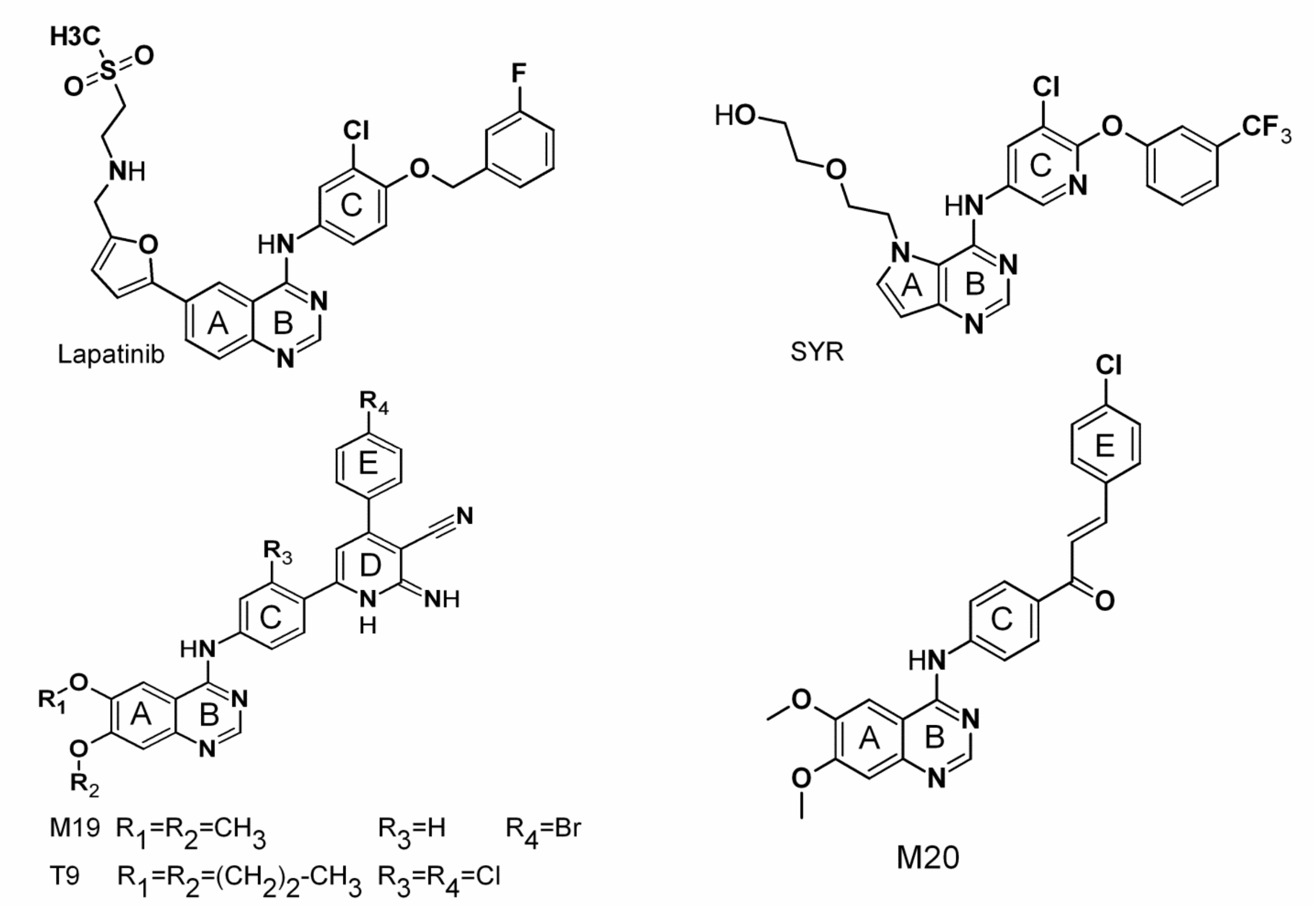

**Figure 2**: Cα backbone atoms RMSDs plots for the studied complex during the 80 ns MD simulations.

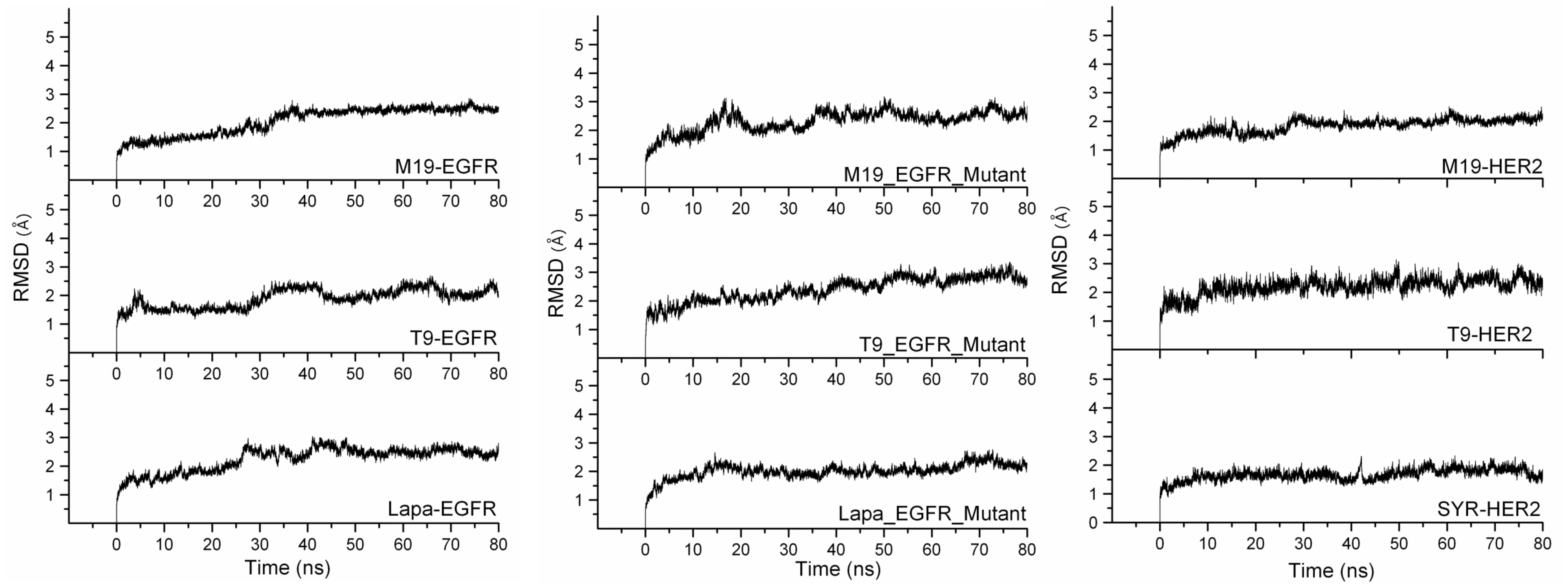

**Figure 3**: Potential binding poses and water orientation of the 9 studied complexes.

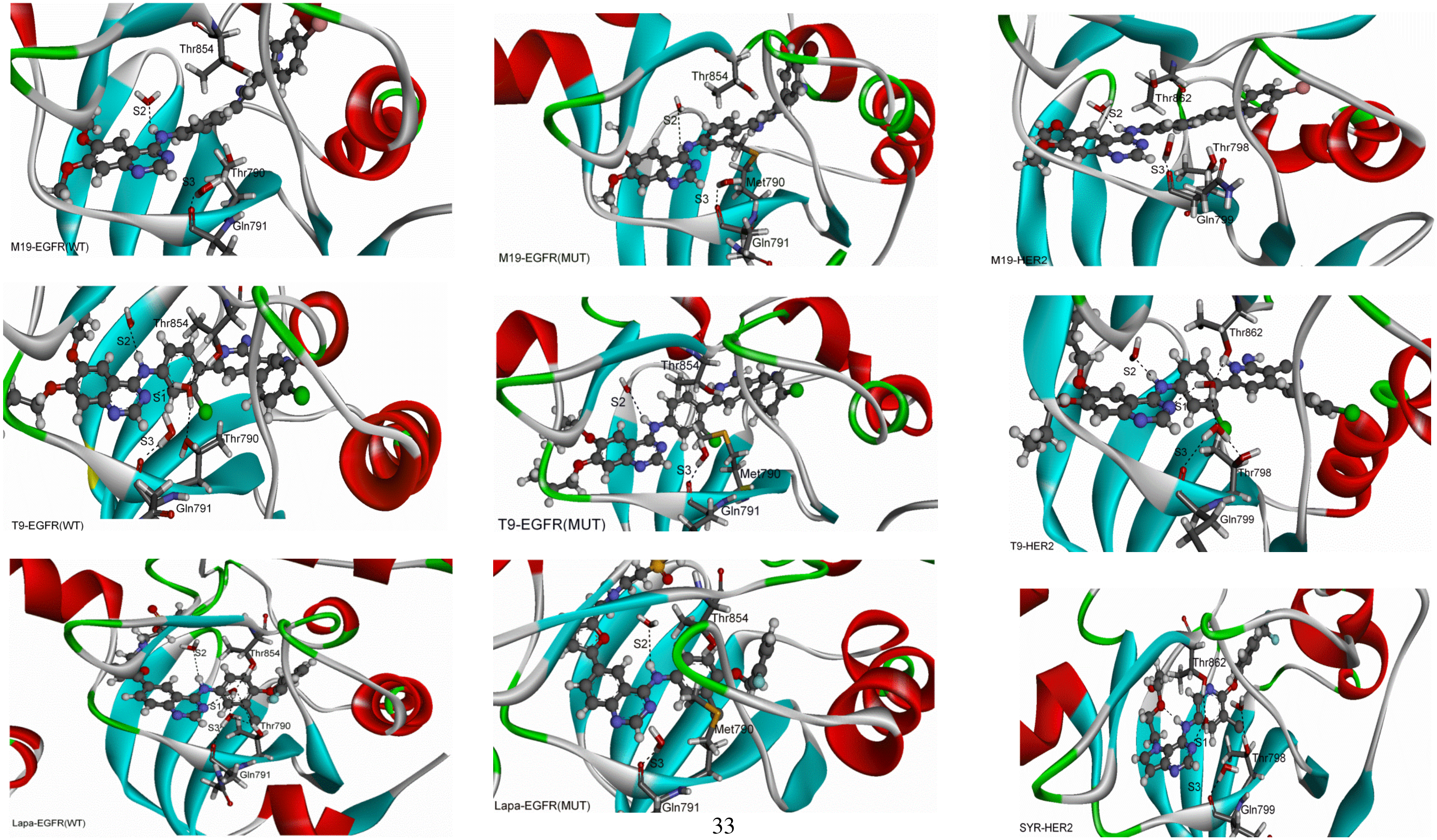

**Figure 4**: Hydrogen bond distance profiles between the inhibitors and the Met hinge residue (Met793 in EGFR and Met801 in HER2) during 80 ns MD simulation.

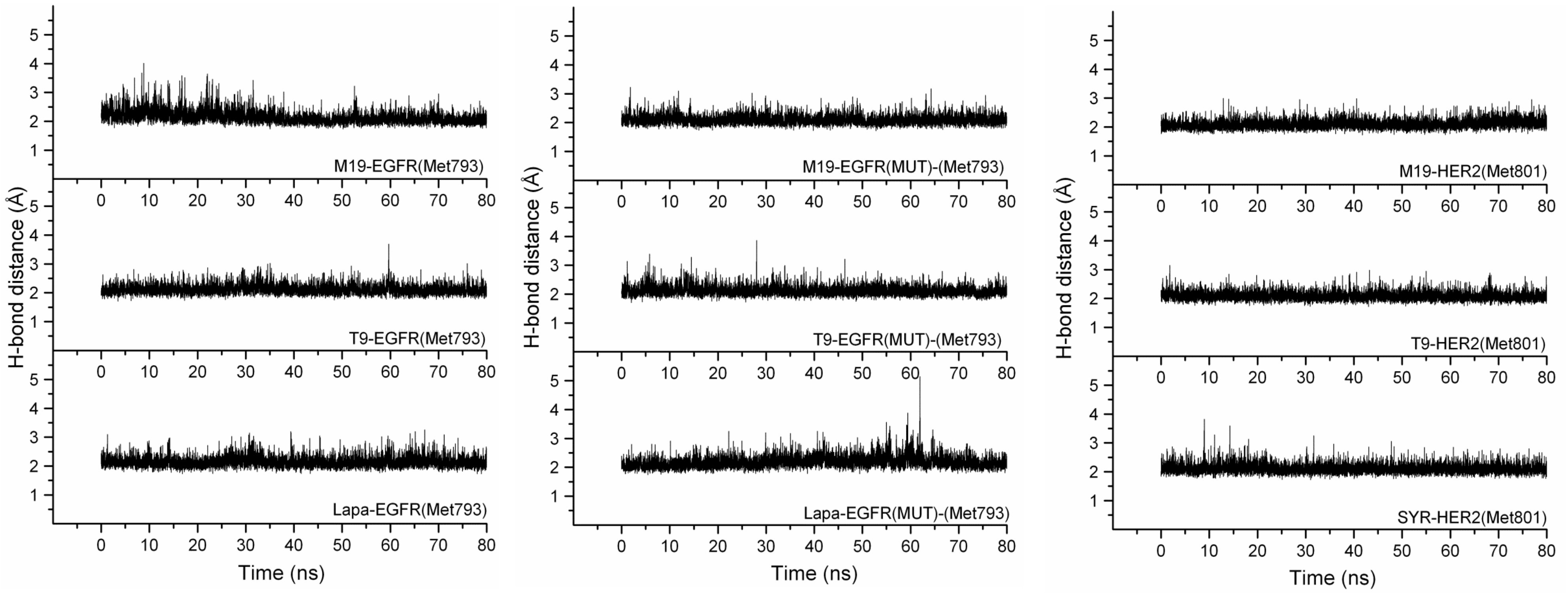

**Figure 5**: 3D interaction with the conserved hinge region Met residue and the additional residues Asp and Lys.

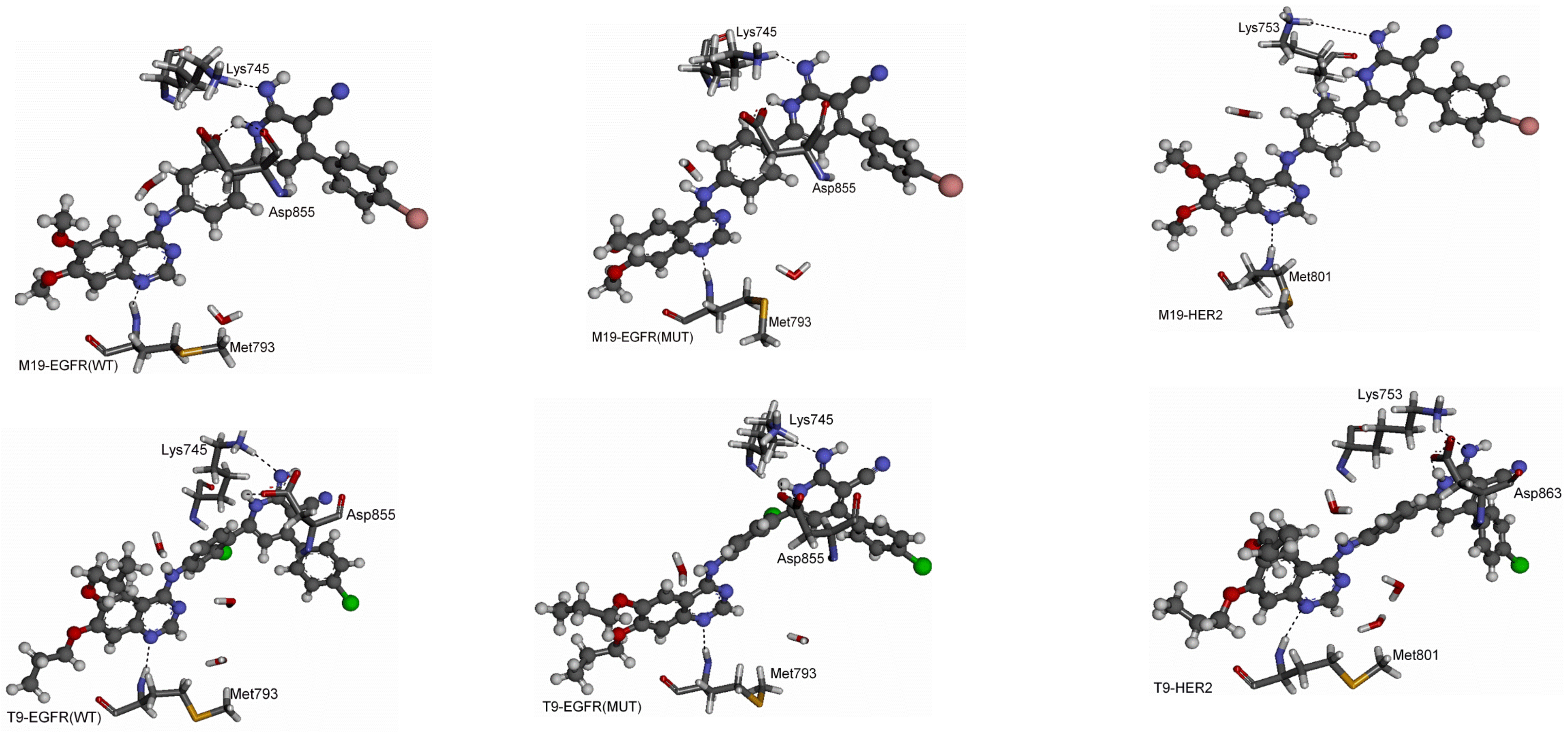

**Figure 6**: Per-residue binding energy decomposition of the 9 MD runs using the MMPBSA module of AMBER

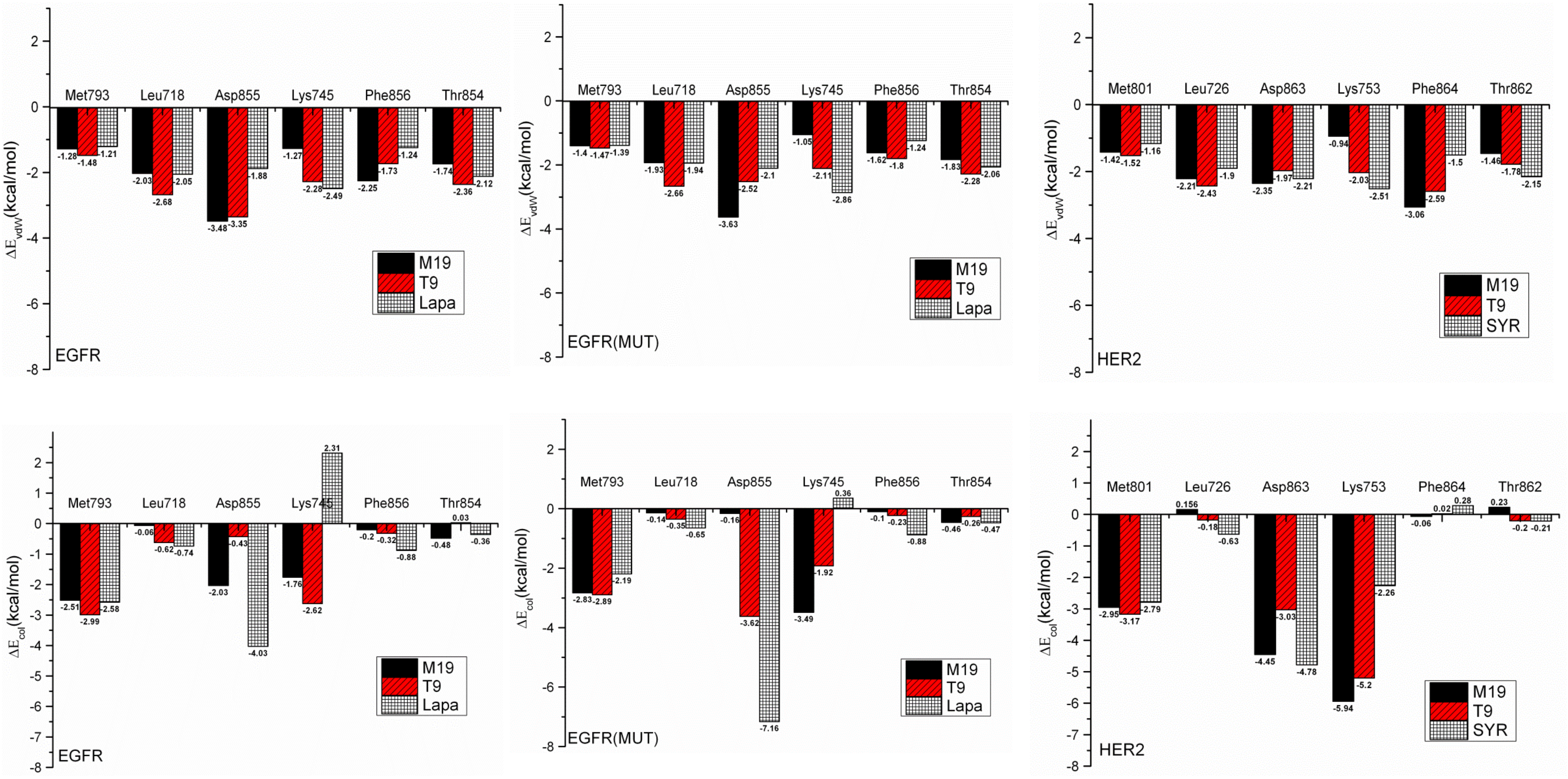